\documentclass{elsart3}
\usepackage{graphicx}
\usepackage{float}
\usepackage{epsfig}
\usepackage{amssymb}
\usepackage{amsmath}
\usepackage{color}
\journal{NIM, Sect. A}

\begin{document}

\begin{frontmatter}
  \title{A survey of energy loss calculations for heavy ions
         \newline between $1$ and $100$ keV}
  
  \author[LIP]{A.~Mangiarotti\thanksref{corauthor}},
  \author[LIP,DPC]{M.~I.~Lopes},
  \author[PIH]{M.~L.~Benabderrahmane},
  \author[LIP,DPC]{V.~Chepel},
  \author[LIP,DPC]{A.~Lindote},
  \author[LIP,DPC]{J.~Pinto~da~Cunha},
  \author[DPF]{P.~Sona}
  
  \thanks[corauthor]{Corresponding author. LIP, Dept. de F\'{\i}sica da
    Universidade de Coimbra, Portugal. Tel.: +351-239-410657; fax:
    +351-239-822358.  {\it E-mail address}: alessio@lipc.fis.uc.pt}
  \address[LIP]{Laborat\'{o}rio de Instrumenta\c{c}\~{a}o e F\'{\i}sica
    Experimental de Part\'{\i}culas (LIP), 3004-516 Coimbra, Portugal}
  \address[DPC]{Departamento de F\'{\i}sica, Universidade de
    Coimbra, 3004-516 Coimbra, Portugal}
  \address[PIH]{Physikalisches Institut der
    Universit\"{a}t Heidelberg, D-69120 Heidelberg, Germany}
  \address[DPF]{Dipartimento di Fisica, Universit\'{a} di Firenze and
    INFN Sez. Firenze, Italy}

\begin{abstract}
  The original Lindhard-Scharff-Schi{\o}tt (LSS) theory and the more
  recent Tilinin theory for calculating the nuclear and electronic
  stopping powers of slow heavy ions are compared with predictions
  from the SRIM code by Ziegler. While little discrepancies are
  present for the nuclear contribution to the energy loss, large
  differences are found in the electronic one.  When full ion recoil
  cascade simulations are tested against the elastic neutron
  scattering data available in the literature, it can be concluded
  that the LSS theory is the more accurate.
\end{abstract}

\begin{keyword}
  Energy loss \sep Stopping power \sep Nuclear recoils \sep Dark matter.
\PACS 34.50.Bw \sep 78.70.-g \sep 95.35.+d.
\end{keyword}

\end{frontmatter}


\section{Introduction}

It is well known that an ion moving inside a medium can loose energy
by collisions with both electrons and nuclei.  Theoretically, it is
necessary to account for both processes to reach an accurate
description of the energy loss below a few keV/amu. From the
experimental point of view, most detectors are just sensitive to
electronic energy loss. Knowledge of the energy loss sharing between
the two processes is mandatory for a detailed understanding of the
response to particles interacting with the detecting medium through
nuclear recoils.  This is the case for Weakly Interacting Massive
Particles (WIMPs), which are possible constituents of the galactic
dark matter.

The purpose of the present work is to explore the different available
descriptions for the two parts of the energy loss between 1 and 100
keV. Range measurements are difficult at such low energies and almost
all published data have been obtained employing elastic neutron
scattering for transferring small and known amounts of energy to atoms
of the detecting material.  To reduce the theoretical difficulties,
the attention is focused on pure substances: hence only symmetric
projectile/target atom combinations will be investigated. From all the
available measurements, known to the authors, the following are then
selected: Si~\cite{Sattler:65,Gerbier:90},
Ge~\cite{Chasman:65,Sattler:66}, and liquid
Xe~\cite{Arneodo:00,Akimov:02,Aprile:05,Chepel:06}. Liquid Ar will
also be considered on account of its interest for dark matter
searches.

Among the few calculations from first principles without free
parameters, there are still the original theory of
Lindhard~\cite{Lindhard:63b} and its reevaluation by
Tilinin~\cite{Tilinin:95}; they will be examined here.

From all the available codes, only SRIM~\cite{Zigler:03} will be
discussed, because: i) it gives separately nuclear and electronic
energy losses, ii) it covers the low energy range of interest, iii)
from an independent survey~\cite{Paul:03} it was found the most
accurate.

\section{The nuclear stopping power}

\begin{figure}[tb]
  \hspace{6mm}\mbox{\mbox{\epsfig{file=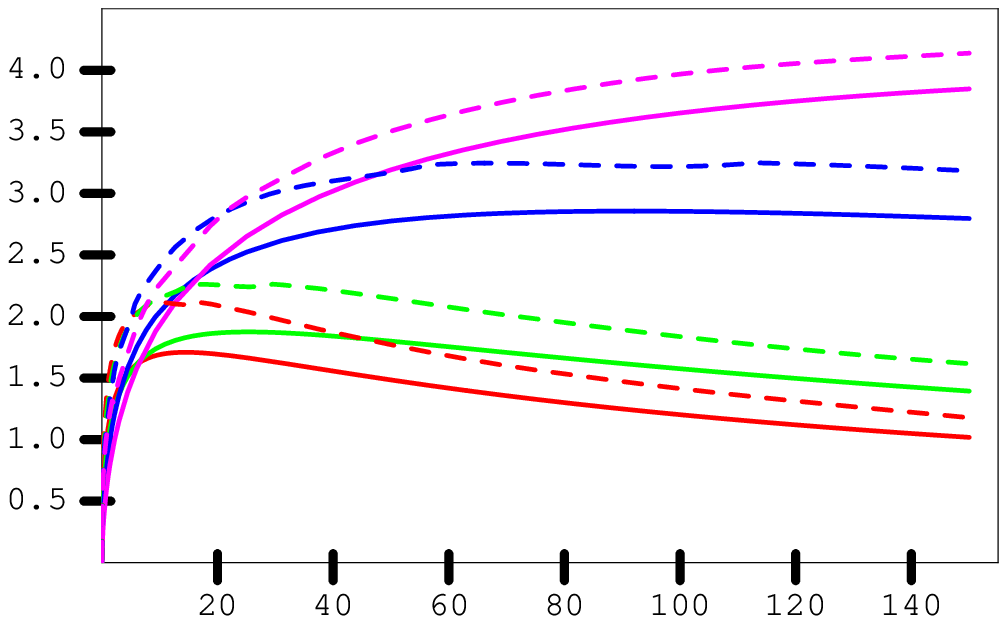,width=.4\textwidth}
      \raisebox{-.02\textwidth}{\hspace{-.22\textwidth} Ion energy
        $E$\ \ \ \ [keV]} \hspace{-.435\textwidth}
      \rotatebox{90}{\hspace{.03\textwidth} $S_n$\ \
        [MeV/(mg/cm$^2$)]}}
    \raisebox{0.065\textwidth}{\hspace{0.065\textwidth} {\small
        \begin{tabular}{rl}
          \rule{1.5mm}{.35mm}\hspace{1mm}\rule{1.5mm}{.35mm} & LSS\\
          \rule{4mm}{.35mm}  & SRIM
        \end{tabular}}}
    \raisebox{0.055\textwidth}{\hspace{0.175\textwidth}\textcolor{red}{Si}}
    \raisebox{0.115\textwidth}{\hspace{-0.026\textwidth}\textcolor{green}{Ar}}
    \raisebox{0.1675\textwidth}{\hspace{-0.031\textwidth}\textcolor{blue}{Ge}}
    \raisebox{0.215\textwidth}{\hspace{-0.033\textwidth}\textcolor{magenta}{Xe}}
  }
  \caption{Nuclear stopping power $S_n$ as a function of the ion
    energy for the symmetric projectile/target combinations considered
    in the present study.}
  \label{S_n}
\end{figure}

According to the Lindhard-Scharff-Schi{\o}tt (LSS)
theory~\cite{Lindhard:63b}, the nuclear stopping power $S_n$ of a
heavy ion is best described by rescaling its energy $E$ and range $R$
to the non-dimensional variables $\epsilon$ and $\rho$, respectively,
defined as
\begin{equation}
\left\{
\begin{aligned}
\epsilon & = C_\mathrm{TF}\,\frac{A_T}{A_\mathrm{tot}}
\,\frac{E/(2\,E_\mathrm{B})}
{Z_P\,Z_T\,Z^{1/2}}\\
\rho & = 4\pi\,(a_\mathrm{B}\,C_\mathrm{TF})^2\;
\frac{A_P\,A_T}{A_\mathrm{tot}^2}
\,\frac{R\,N}{Z} 
\end{aligned}
\right.\;,
\label{LSS_scal}
\end{equation}
with $Z=Z_P^{2/3}+Z_T^{2/3}$ and $A_\mathrm{tot}=A_P+A_T$. In
Eq.~(\ref{LSS_scal}) $N$ is the number density of the target material,
$Z_P$ and $A_P$ are the atomic and atomic mass numbers of the
projectile atom, respectively, $Z_T$ and $A_T$ are the correspondents
for the target atom, $a_\mathrm{B}$ is the Bohr radius, $E_\mathrm{B}$
the Bohr energy and $C_\mathrm{TF}$ the Thomas-Fermi constant
$(9\pi^2/2^7)^{1/3}$. In fact, while the nuclear part of the energy
loss $(dE/dx)_n$ depends on the projectile and target ions,
$(d\epsilon/d\rho)_n$ depends on the Thomas-Fermi interaction
potential alone and is a universal function $f(\epsilon)$, which can
be calculated numerically~\cite{Lindhard:68}. The values of $S_n$ are
reproduced with the physical units in Fig.~\ref{S_n}.

A similar rescaling is employed in SRIM ($Z$ in all appearances in
Eq.~(\ref{LSS_scal}) is replaced by an empirical
$Z^{1/2}=Z_P^{0.23}+Z_T^{0.23}$), but now the universal function
$f(\epsilon)$ is determined not from first principles but from a fit
to experimental data. The results are also shown in Fig.~\ref{S_n}.

While differences are small, it is known from sputtering
data~\cite{Oetzmann:75} that the LSS theory overestimates $S_n$. This
problem is reduced in SRIM, as it is based on data themselves.
Lindhard et al.~\cite{Lindhard:68} had also warned that the
Thomas-Fermi treatment might be inadequate for very low energies
(i.e.  $\epsilon<10^{-2}$), where mostly the tails of the ion-ion
potential are probed. In the present study, this is of concern only for
Xe nuclear recoils below $15$ keV.

\section{The electronic stopping power}

\begin{figure}[tb]
  \hspace{6mm}\mbox{\mbox{\epsfig{file=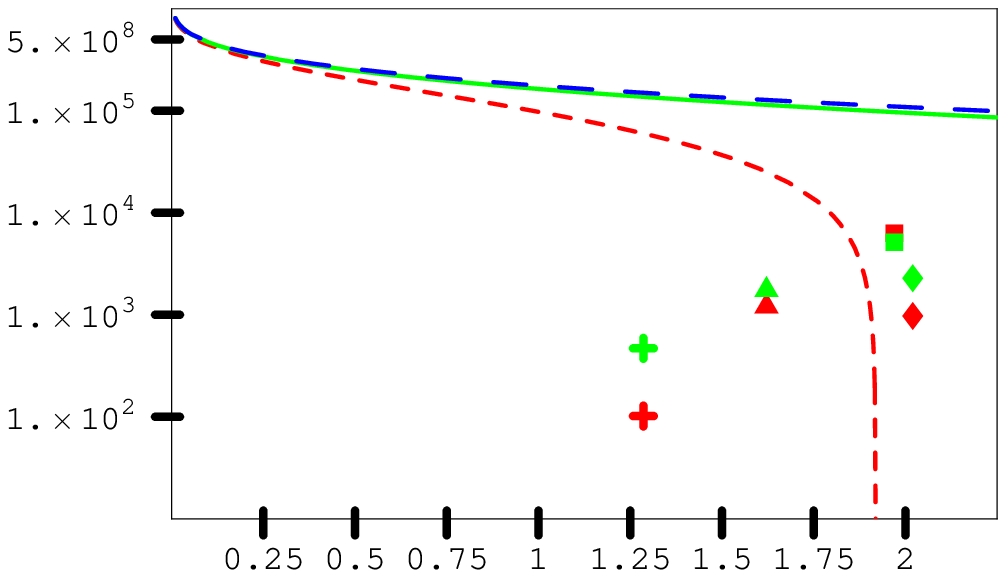,width=.4\textwidth}
      \raisebox{-.0125\textwidth} {\hspace{-.298\textwidth} Wigner-Seitz radius $r_s$\ \ \ \ [$a_\mathrm{B}$]}
      \hspace{-.437\textwidth} \rotatebox{90}{\hspace{-.04\textwidth}
        $(dE/dx)_e/(\beta\,Z_P^2)$\ \ [MeV/cm]}}
    \raisebox{0.1125\textwidth}{\hspace{0.08\textwidth} {\small
        \begin{tabular}{rl}
          \textcolor{red}{\rule{1mm}{.35mm}\hspace{1mm}\rule{1mm}{.35mm}\hspace{1mm}\rule{1mm}{.35mm}\hspace{1mm}\rule{1mm}{.35mm}} & Fermi-Teller\\
          \textcolor{blue}{\rule{2.5mm}{.35mm}\hspace{2mm}\rule{2.5mm}{.35mm}}  & Lindhard-Winther\\
          \textcolor{green}{\rule{7mm}{.35mm}} & Ritchie \\
          \textcolor{green}{$\bullet\;\;\;$}   & LSS\\
          \textcolor{red}{$\bullet\;\;\;$}     & SRIM
        \end{tabular}}}
    \raisebox{0.1575\textwidth}{\hspace{0.059\textwidth}Si}
    \raisebox{0.115\textwidth}{\hspace{-0.059\textwidth}Ar}
    \raisebox{0.0875\textwidth}{\hspace{0.01\textwidth}Ge}
    \raisebox{0.08\textwidth}{\hspace{-0.125\textwidth}Xe}
  }
  \vspace{1mm}
\caption{Proportionality coefficient of the electronic energy loss to
  the particle velocity as a function of the Wigner-Seitz radius $r_s$.
  For ions a $Z_P^2$ scaling is applied.}
\label{S_e}
\end{figure}

The electronic energy loss $(dE/dx)_e$ of a unitary charge particle
with a velocity $\beta$ was described as an interaction with an
electron plasma in the original work of Fermi and Teller.  They
explicitly distinguished two cases, for $\beta$ above and below the
Fermi velocity $\beta_F$. For $\beta<\beta_F$, $(dE/dx)_e$ was found
to be proportional to $\beta$, with a proportionality coefficient
being a unique function of the electron plasma density $n_0$, usually
expressed in terms of the Wigner-Seitz radius $r_s=(3
n_0/(4\pi))^{1/3}$~\cite{Ferrell:77}.  Their result is plotted in
Fig.~\ref{S_e}. Typically, $r_s$ needs to be corrected because tightly
bound electrons contribute only marginally to $(dE/dx)_e$ much below
the Bragg peak.  In crystals, like Si and Ge, this effective density
of the free electron plasma can be deduced from optical
properties~\cite{Mann:81}.  In liquids, like Ar and Xe, the problem is
much more difficult and no correction was attempted in Fig.~\ref{S_e}.
Successively, Lindhard calculated in a self consistent way the local
increase of the electron plasma density around the intruder particle
due to its Coulomb field.  This leads to a higher $(dE/dx)_e$ as can
be seen in Fig.~\ref{S_e}.  Finally, Ritchie considered the case where
the Coulomb field is exponentially screened, slightly decreasing
$(dE/dx)_e$ (see Fig.~\ref{S_e}). The last effect is of particular
relevance for ions, which can accommodate bound states while sweeping
through the electron plasma. For a bare ion, a scaling with $Z_P^2$ to
the elementary particle case is expected, as assumed in
Fig.~\ref{S_e}. In reality, this is not correct and the theory was
extended to a partially ionized intruder by Ferrell and
Ritchie~\cite{Ferrell:77}, but the determination of the equilibrium
charge of a given ion remains a difficult task.  Lindhard also
independently investigated this problem~\cite{Lindhard:63b} and, using
the Thomas-Fermi theory, arrived at a closed form for the
proportionality coefficient. In terms of the non-dimensional variables
introduced in Eq.~(\ref{LSS_scal}), his result can be expressed as
$(d\epsilon/d\rho)_e=\kappa\,\sqrt{\epsilon}$ where
\begin{equation}
\kappa=\frac{32}{3\,\pi}\,
\sqrt{\frac{m_e\,c^2}{m_\mathrm{amu}\,c^2}}\,
\frac{Z_P^{1/2}\,Z_T^{1/2}}{Z^{3/4}}\,
\frac{A_\mathrm{tot}^{3/2}}{A_P^{3/2}\,A_T^{1/2}}\;
\xi_e
\label{kappa}
\end{equation}
with $\xi_e\approx Z_p^{1/6}$ (which is regarded only as an
approximation by Lindhard)~\cite{Lindhard:63b}.  The points
corresponding to the projectile/target combinations of interest for
the present study are also reported in Fig.~\ref{S_e}. The suppression
of the electronic energy loss occurs mostly due to the partial
ionization of the intruder. It strongly increases with $Z_P$
($Z_P=Z_T$).

The proportionality of $(dE/dx)_e$ with $\beta$ is also a feature of
the SRIM code, allowing a value of $(dE/dx)_e/(\beta\,Z_P^2)$ to be
extracted (see Fig.~\ref{S_e}). While SRIM exceeds the LSS theory for
Si, it then decreases consistently below it, up to a factor of $4$ for
Xe.  The reason for this discrepancy is unclear, because the details
on the implementation of $(dE/dx)_e$ in SRIM for low velocities are
not public, but it probably resides in the estimate of the intruder
charge state. It has been verified that data for protons of comparable
energies per nucleon on Ar and Xe are well reproduced by SRIM.

The big drawback of the described theoretical approaches is to assume
that the electronic and nuclear collisions are
uncorrelated~\cite{Lindhard:63b}. In reality, the screened Coulomb
repulsion between the two interacting nuclei makes part of the impact
parameter range unavailable for the scattering of the electrons
belonging to the target atom in the screened Coulomb field of the
projectile and vice versa. Tilinin~\cite{Tilinin:95} has shown that
the final net effect is a great decrease of $S_e$ for $\epsilon \ll
1$, with a corresponding lack of proportionality to $\sqrt{\epsilon}$.
His results can also be recast in the form of Eq.~(\ref{kappa}) where
$\xi_e$ is replaced by a function $\tau(\epsilon,Z_P/Z_T)$ that can be
tabulated~\cite{Tilinin:95}. In the present case, his theory predicts
roughly half the value of $S_e$ expected from LSS and SRIM for Si and
approximately agrees with SRIM for Xe.

\section{The full ion recoil cascade}

\begin{figure}[tb]
  \hspace{7mm}\mbox{\parbox{0.375\textwidth}{
      \mbox{\epsfig{file=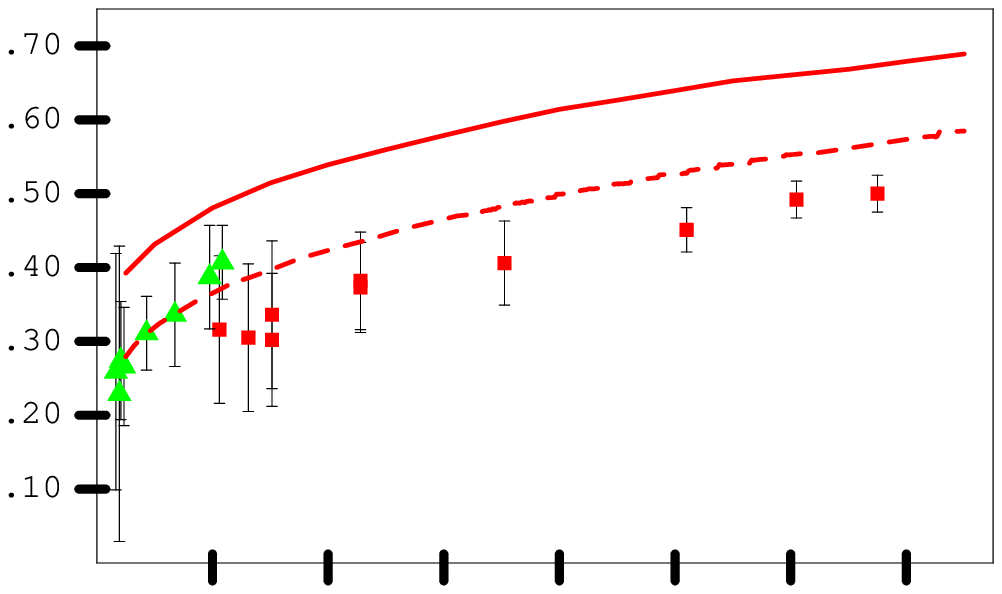,width=.375\textwidth}
        \raisebox{0.05\textwidth}{\hspace{-0.175\textwidth} {Si\ \ \tiny
            \begin{tabular}{rl}
              \textcolor{red}{$\blacksquare$}     & Ref.~\cite{Sattler:65}\\
              \textcolor{green}{$\blacktriangle$} & Ref.~\cite{Gerbier:90}
            \end{tabular}}}
        \raisebox{0.05\textwidth}{\hspace{-0.24\textwidth} {\small
            \begin{tabular}{rl}
              \textcolor{red}{\rule{1.5mm}{.35mm}\hspace{1mm}\rule{1.5mm}{.35mm}} & LSS\\
              \textcolor{red}{\rule{4mm}{.35mm}} & TRIM
            \end{tabular}}}
      }\vspace{-3.9mm}
      \mbox{\epsfig{file=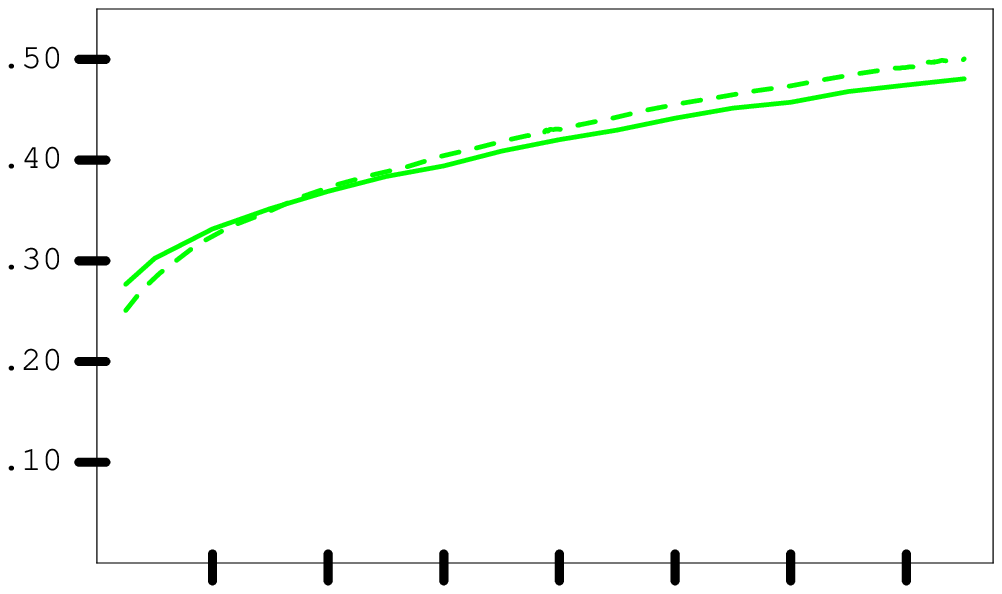,width=.375\textwidth}
        \raisebox{0.05\textwidth}{\hspace{-0.175\textwidth} {Ar}}
        \raisebox{0.05\textwidth}{\hspace{-0.15\textwidth} {\small
            \begin{tabular}{rl}
              \textcolor{green}{\rule{1.5mm}{.35mm}\hspace{1mm}\rule{1.5mm}{.35mm}} & LSS\\
              \textcolor{green}{\rule{4mm}{.35mm}} & TRIM
            \end{tabular}}}
      }\vspace{-3.9mm}
      \mbox{\epsfig{file=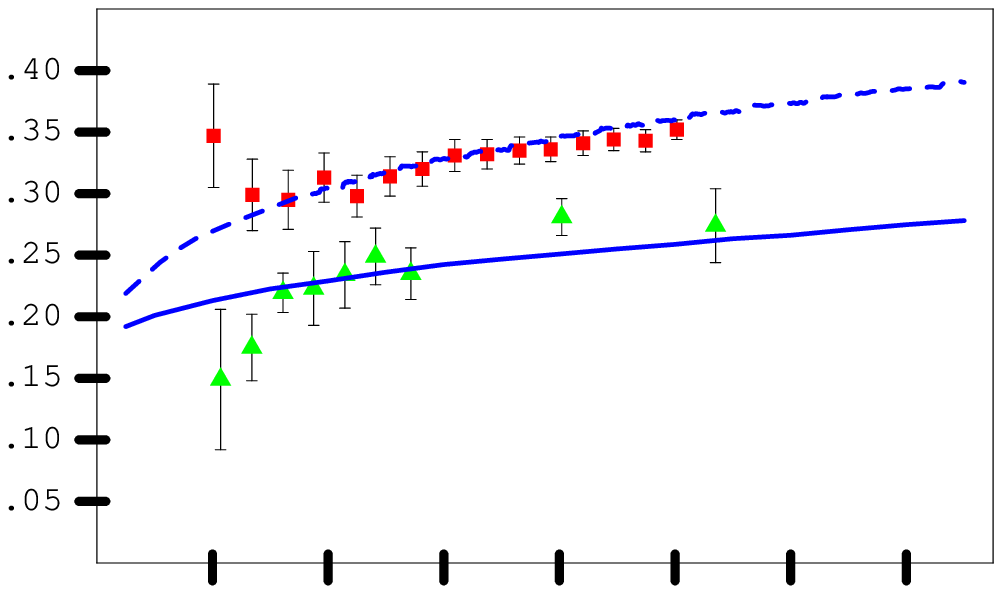,width=.375\textwidth}
        \raisebox{0.05\textwidth}{\hspace{-0.175\textwidth} {Ge\ \ \tiny
            \begin{tabular}{rl}
              \textcolor{red}{$\blacksquare$}      & Ref.~\cite{Chasman:65}\\
              \textcolor{green}{$\blacktriangle$}  & Ref.~\cite{Sattler:66}
            \end{tabular}}}
        \raisebox{0.05\textwidth}{\hspace{-0.24\textwidth} {\small
            \begin{tabular}{rl}
              \textcolor{blue}{\rule{1.5mm}{.35mm}\hspace{1mm}\rule{1.5mm}{.35mm}} & LSS\\
              \textcolor{blue}{\rule{4mm}{.35mm}}  & TRIM
            \end{tabular}}}
      }\vspace{-2.7mm}
      \mbox{\epsfig{file=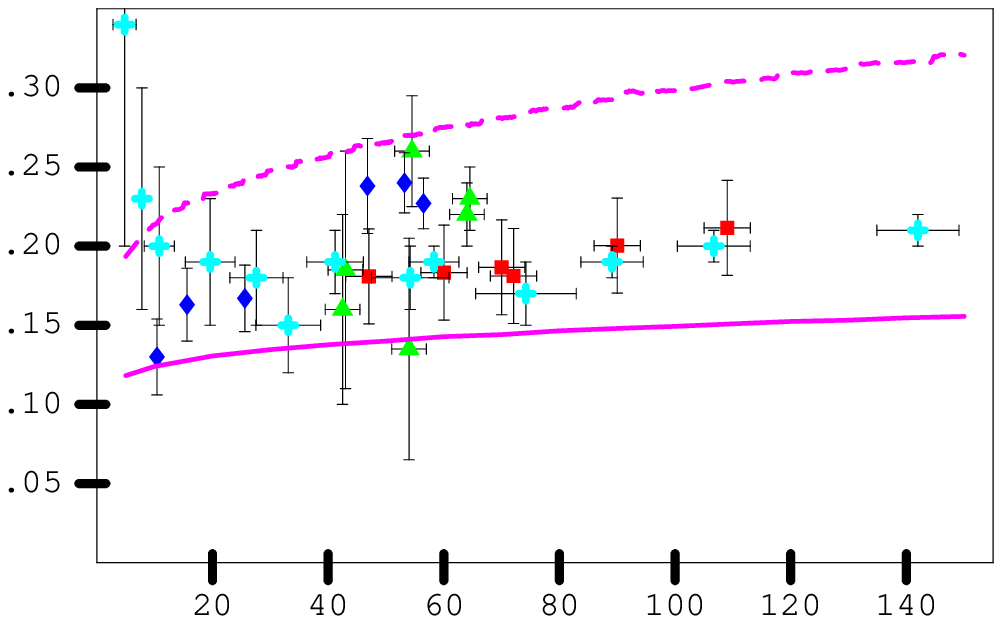,width=.375\textwidth}
        \raisebox{0.07\textwidth}{\hspace{-0.175\textwidth} {Xe\ \ \tiny
            \begin{tabular}{rl}
              \textcolor{red}{$\blacksquare$}      & Ref.~\cite{Arneodo:00}\\
              \textcolor{green}{$\blacktriangle$}  & Ref.~\cite{Akimov:02}\\
              \textcolor{blue}{$\blacklozenge$}    & Ref.~\cite{Aprile:05}\\
              \textcolor{cyan}{{\bf +}}            & Ref.~\cite{Chepel:06}
            \end{tabular}}}}
      \raisebox{0.055\textwidth}{\hspace{-0.24\textwidth} {\small
          \begin{tabular}{rl}
            \textcolor{magenta}{\rule{1.5mm}{.35mm}\hspace{1mm}\rule{1.5mm}{.35mm}} & LSS\\
            \textcolor{magenta}{\rule{4mm}{.35mm}} & TRIM
          \end{tabular}}}
    }\raisebox{-.45\textwidth}
    {\hspace{-.214\textwidth} Ion energy $E$\ \ \ \ [keV]}
    \hspace{-.4\textwidth}\rotatebox{90}
    {\hspace{-.22\textwidth} Total fraction of energy transferred to electrons}}
\caption{Total fraction of the initial ion energy transferred to electrons
 integrated over the full cascade as a function of the ion energy itself.}
\label{fullcasc}
\end{figure}

As mentioned, resort must be made to response measurements employing
elastic neutron scattering, where recombination or quenching may
influence the final fraction of the total energy transferred to
electrons that is detectable as excitation or ionization, particularly
in the case of scintillation yield for LXe.  Theoretically, however,
an even bigger disadvantage is present; especially for high $Z$
elements: $S_n$ dominates over $S_e$ and most of the primary ion
energy is transferred to nuclear recoils.  The knocked ion undergoes
the same process, resulting in a full cascade of recoils, whose total
electronic energy loss must be evaluated. In the case of the SRIM
code, a second program, called TRIM~\cite{Zigler:03}, reads $S_n$ and
$S_e$ from the first and performs the computation.  Lindhard and his
group solved numerically the transport equations corresponding to the
LSS theory and arrived at a parameterization in terms of $\epsilon$
and $\kappa$ (see Eq.~(9) of Ref.~\cite{Lindhard:63a}).  Both results
are compared with data in Fig.~\ref{fullcasc} (the two series of
points for Ge are inconsistent). The LSS theory is on average better,
while SRIM both overpredicts and strongly underpredicts the data for
Si and Xe, respectively. While the first discrepancy could be ascribed
to recombination, the second appears even more surprising due to the
probable presence of quenching.

\section{Conclusions}

For slow heavy ions, the nuclear stopping power predicted by the
original LSS theory and the current SRIM code differ at most by
$\approx 15\%$. On the contrary, for the electronic stopping power,
big discrepancies are present between the LSS theory, the theory of
Tilinin and SRIM (up to a factor of $\approx 4$ for Xe). Judging from
the elastic neutron scattering data, the LSS theory seems the best of
all. More detailed full cascade simulations will be performed in the
future to assess the robustness of this conclusion. New experimental
data for Germanium would be highly needed for clarifying the
situation.

\section*{Acknowledgments}
This work was supported by FCT/\-FEDER/\-POCI-2010 fund (project
POCI/FP/63446/2005).

%



\begin{thebibliography}{99}
\bibitem{Lindhard:63b} J.~Lindhard et al., Mat. Fis. Medd. Dan. Vid. Selsk.
                       33 (1963) No.14.
\bibitem{Tilinin:95} I.S.~Tilinin, Phys. Rev. A 51 (1995) 3058.
\bibitem{Sattler:65} A.R.~Sattler, Phys. Rev. 138 (1965) A1815.
\bibitem{Gerbier:90} G.~Gerbier et al., Phys. Rev. D 42 (1990) 3211.
\bibitem{Chasman:65} C.~Chasman et al., Phys.  Rev. Lett. 15 (1965) 245.
\bibitem{Sattler:66} A.R.~Sattler et al., Phys. Rev. 143 (1966) 588.
\bibitem{Arneodo:00} F.~Arneodo et al., Nucl. Instr. Meth. A 449 (2000) 147.
\bibitem{Akimov:02} D.~Akimov et al., Phys. Lett. B 524 (2002) 245.
\bibitem{Aprile:05} E.~Aprile et al., Phys. Rev. D 72 (2005) 072006.
\bibitem{Chepel:06} V.~Chepel et al., Astrop. Phys. 26, (2006) 58.
\bibitem{Zigler:03} J.F.~Zigler, Program SRIM/TRIM, version 2003.26,
                    obtained from http://www.srim.org.
\bibitem{Paul:03} H.~Paul et al., Nucl. Instr. Meth. B 209 (2003) 252.
\bibitem{Lindhard:68} J.~Lindhard et al., Mat. Fis. Medd. Dan. Vid. Selsk.
                      36 (1968) No.10.
\bibitem{Oetzmann:75} H.~Oetzmann et al., Phys. Lett. A 55 (1975) 170.
\bibitem{Ferrell:77} T.L.~Ferrell et al., Phys. Rev. B 16 (1977) 115.
\bibitem{Mann:81} A.~Mann et al., Phys. Rev. B 9 (1981) 4999.
\bibitem{Lindhard:63a} J.~Lindhard et al., Mat. Fis. Medd. Dan. Vid. Selsk.
                       33 (1963) No.10.
\end{thebibliography}
\end{document}